\documentclass{aa}

\usepackage{graphicx}
\usepackage{xcolor}
\usepackage{bm}
\usepackage{mathtools}

\usepackage{txfonts}

\usepackage{hyperref}
\hypersetup{
  colorlinks=true,
  breaklinks=true,
  allcolors=black,
}

\begin{document}

   \title{The intracluster light as an estimator of the cluster mass profile}

   \author{I. Alonso Asensio
          \inst{1,2}
          \and
          A. Contreras-Santos
          \inst{1,2}
          }

   \institute{Instituto de Astrof\'isica de Canarias, E-38205, La Laguna, Tenerife, Spain
         \and
             Departamento de Astrof\'isica, Universidad de La Laguna, E-38206, La Laguna, Tenerife, Spain
             }

   \date{Received -; accepted -}

  \abstract
   {The intracluster light (ICL) comprises stars that are not bound to individual galaxies within a galaxy cluster, and it provides insights into the cluster mass distribution, evolutionary history, and dynamical state.}
   {We study the viability of the intracluster stellar mass as a proxy for computing the total mass profiles of galaxy clusters.}
   {High-resolution simulations from the C-EAGLE project were used to study the ratio of the intracluster stellar mass and total matter projected densities. This ratio follows a power law, and we present a model for its fit parameters and associated errors.}
   {We used this relation to estimate the mass profile of the Perseus cluster based on Euclid observations that extend up to one-third of the virial radius. The obtained cluster mass is compatible with other measurements from galaxy velocity dispersion, but it is overestimated by a factor of two compared to X-ray mass estimates. We repeated this process for four clusters in the Hubble Frontier Fields, finding compatibility with weak- and strong-lensing mass estimates.}
   {This method provides an independent approach to cluster mass estimation that is based solely on the observed ICL and a simulation-calibrated relation.}

   \maketitle

\section{Introduction}

The intracluster light (ICL) consists of stars that are not gravitationally bound to individual galaxies, but are bound in a galaxy cluster. They are observed as a diffuse distribution of stars in the cluster. These stars are typically stripped from their host galaxies through mergers and tidal interactions, and their spatial distribution can extend to the outer regions of the cluster. As a result, the ICL can provide important information about the total mass distribution, evolutionary history, and dynamical state of the cluster \citep[e.g.,][respectively]{Montes19, Mihos05, JimenezTeja18}.

It has been widely studied whether the fraction of the stellar mass in ICL ($f_\text{ICL}$) correlates with the total mass of the cluster. However, there is no clear consensus about the dependence of $f_\text{ICL}$ on cluster mass (see \citealp{Contini2021, Montes2022a} for reviews, and \citealp{Ragusa2023,Contreras-Santos2024,Zhang2024,Montenegro-Taborda2025}, for recent works on the topic). This relation can be affected by the exact definition of the ICL \citep{Cui2014, Kluge2021}. Thus, further exploration of the ICL use in estimating cluster masses has been limited.

With current observational datasets, rather than just examining $f_\text{ICL}$, the mass profile of the ICL can be measured. This can be achieved via stacking \citep{Zhang2024} or directly from deep imaging of clusters. Some notable examples of the latter are the ICL profile obtained by \citet{Montes2014} and then extended by \citet{Montes18} to the Frontier Fields Clusters observed with the Hubble Space Telescope (HST) \citep{Lotz2017a}. These works allowed for a detailed study of the ICL spatial distribution and its correlation with the mass distribution obtained from weak lensing \citep{Montes19}. Together, these studies helped to establish the concept that the ICL can be used to trace the underlying dark matter potential, and later works explored this further from both theoretical \citep{AlonsoAsensio2020, Contini2020, Gonzalez2021, Shin2022, Yoo2024, Contreras-Santos2024, Bilata-Woldeyes2025} and observational \citep{Sampaio-Santos2020, Yoo2021, Diego2023a} points of view.

Of these follow-up studies, we highlight the study by \citet{AlonsoAsensio2020}. This study used the C-EAGLE set of simulations to quantitatively compare the similarity between the intracluster stars and total matter projected densities. The authors reported that the similarity was greater than observed, which reinforced the hypothesis that the intracluster stellar distribution can be used to infer the total matter distribution. \citet{AlonsoAsensio2020} further proposed a simulation-calibrated method for estimating the mass of a cluster from the observed ICL profile based on a power-law relation between the projected stellar mass and the total mass density profiles.

A recent work that again focused on the ICL as a mass tracer is the early release from Euclid, which reported observations of the Perseus cluster \citep{Kluge2025}. The ICL was measured up to a radius of 600 kpc, which is equivalent to one-third of the cluster virial radius. This work provides an exquisite dataset to explore how deep observations of the ICL can be used to estimate the cluster mass following the idea first presented by \citet{AlonsoAsensio2020}.

In this work, we aim to extend and better quantify the relation found by \citet{AlonsoAsensio2020}, with particular emphasis on the realistic evaluation of the associated errors. We apply it to observational data to obtain cluster mass profiles. We make use of the Perseus cluster observations with Euclid to showcase the ideal scenario in which the ICL profile can be observed up to the cluster outskirts and the mass profile can be fully recovered. We also apply the method to less optimal cases in which the radial extent of the ICL is smaller and some extrapolation must be performed.

The paper is structured as follows. In Sect.~\ref{sec:sims} we revisit the relation between ICL and total mass profiles by \citet{AlonsoAsensio2020} and extend it to evaluate realistic errors. In Sect.~\ref{sec:methods} we apply the relation to estimate the mass profiles of observed clusters. Applications of this method to Euclid and HST data are shown in Sect.~\ref{sec:results}. The results are discussed in detail in Sect.~\ref{sec:discussion}, and the concluding remarks are laid out in Sect.~\ref{sec:conclusions}.

\section{The ICL-total mass relation in simulations} \label{sec:sims}

\citet{AlonsoAsensio2020} showed that the ratio of the projected intracluster stellar mass and the total mass profiles of simulated clusters follows a power law (their Equation 4 and 5). This relation indicates that the relative amount of intracluster stellar mass decreases more rapidly toward larger radii than the total mass. \citet{AlonsoAsensio2020} provided a simple fit to the mean profile of the 30 C-EAGLE clusters. However, this fit produced very small errors that did not accurately represent the true scatter in the studied ratio.

Our goal is to use this relation to estimate the mass of clusters from ICL observations, which requires a realistic error propagation. To expand on the work of \citet{AlonsoAsensio2020}, we reanalyzed the simulations incorporated more projections, and provided realistic scatter estimates.

As \citet{AlonsoAsensio2020}, we used the set of 30 zoom-in cluster simulations performed within the C-EAGLE project \citep{Barnes2017,Bahe2017}. The clusters are uniformly distributed in the $[10^{14}, 10^{15.4}]\, \text{M}_{200}/\text{M}_\odot$ mass range, where M$_{200}$ is the halo mass.\footnote{M$_{200}$ is the mass enclosed in a sphere of radius $r_{200}$ whose mean density equals 200 times the critical density of the Universe.}
The simulations are performed using the EAGLE model for galaxy formation and evolution \citep{Schaye2015}, with the AGNdT9 calibration. The simulations have a spatial resolution of $\epsilon = 0.7$ kpc and a baryonic mass resolution of $m_\text{gas} \approx 1.81 \times 10^6\,\text{M}_\odot$.
The reader is referred to the EAGLE model description \citep{Schaye2015} and calibration \citep{Crain2015} for further details.

For the analysis, we defined intracluster particles as those belonging to the main halo of the largest friends-of-friends group as computed by SUBFIND \citep{Dolag2009}. This excluded particles that are gravitationally bounded to satellite galaxies, but included those associated with the brightest central galaxy (BCG).
Throughout the paper, we refer to the stellar component of the intracluster as ICL+BCG to ensure consistency with the terminology that is commonly adopted in the literature \citep[e.g.,][]{Kluge2021,Yoo2024,Kluge2025}.
For each cluster, we computed the projected radial density profiles of the intracluster stellar mass ($\Sigma_\star$) and the total mass ($\Sigma_\text{tot}$), centered on the halo center for three independent projections.
Deviations from circularity were not taken into account, but we experimented with ellipse fitting and found no significant changes in the conclusions.

In the last step, we computed $\Sigma_\star / \Sigma_\text{tot}$ and fit it with a power law from 10 kpc to $r_{200}$,\footnote{We experimented with different outer limits for the fit, but found no choice that presented better results. For the inner limit, we chose a value large enough to avoid the center of the cluster, where $\Sigma_\star / \Sigma_\text{tot} \lesssim 1$.}
\begin{equation}
  \log_{10} \Sigma_\text{tot}[\text{M}_\odot \text{kpc}^{-2} ] = \log_{10} \Sigma_\star[\text{M}_\odot \text{kpc}^{-2}] - a \log_{10} (r [\text{kpc}]) - b
    \label{eq:fit}.
\end{equation}

\begin{figure}
  \resizebox{\hsize}{!}{\includegraphics{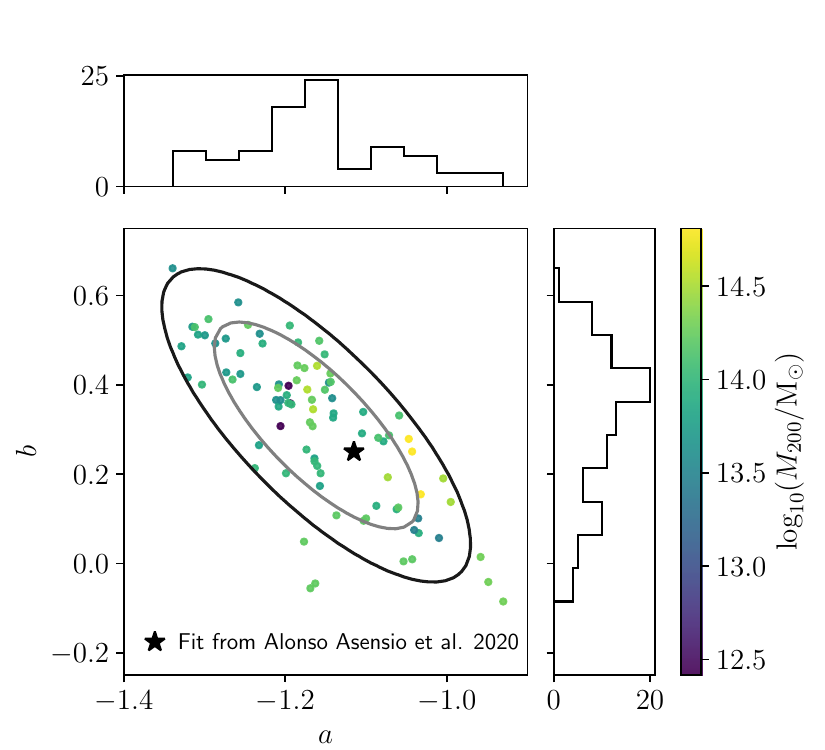}}
  \caption{Corner plot of the fit parameters of equation \ref{eq:fit}. The contours show the 1 and 2$\sigma$ regions of the multivariate normal distribution of the fits.}
  \label{fig:parameters}
\end{figure}

For each epoch ($z=0, 0.1, 0.2, 0.35, 0.54$), 90 values of $a$ and $b$ were obtained (three projections per cluster). To obtain an error estimate that included the correlation between the parameters, we fit a multivariate normal distribution to the set of $(a,b)$ points. The corner plot for the fit parameters at $z=0$ is shown in Fig.~\ref{fig:parameters}. The color scale represents the mass of the clusters.
Within the studied mass range, there is no strong correlation with the mass of the cluster (see also Appendix~\ref{app:mass}). The contours are drawn at the 1 and 2$\sigma$ deviations of the resulting multivariate normal distributions. As reference, the best-fit from \citet{AlonsoAsensio2020} is shown as a star.
Although it lies within the 1$\sigma$ region, the error estimate given by \citet{AlonsoAsensio2020} was  $\pm 0.005$ and $\pm 0.01$ for $a$ and $b$, respectively, which highly underestimates the scatter shown in Fig.~\ref{fig:parameters}.

We provide a table with the fit parameters to multivariate normal distributions for each epoch in Appendix~\ref{app:fit_parameters}.
Hereafter, unless otherwise stated, we use the fit given by joining all the data with $z<0.5$,

\begin{align*}
  \bar{a} &= -1.139,\; &   \bar{b} &= 0.316,\;   & \\
      \Sigma_a &= 0.00958,\;  & \Sigma_b &= 0.036,\;  &  \Sigma_{ab} = -0.0146;
\end{align*}

where $\bar{a}$ and $\bar{b}$ are the mean values of the power-law fit
(Equation~\ref{eq:fit}). $\Sigma$ is the covariance matrix. Its diagonal members are
the variances $\Sigma_a$ and $\Sigma_b$, and the off-diagonal value
$\Sigma_{ab}=\Sigma_{ba}$ is the covariance.

\section{Applying the ICL-total mass relation to observations} \label{sec:methods}

Given the relation between ICL and total mass (Equation~\ref{eq:fit}) and the fit obtained from simulated clusters (Appendix~\ref{app:fit_parameters}), the mass profile of a cluster can be derived from its ICL profile.

In observations, the ICL profile is typically measured up to a few hundred kiloparsecs due to its low surface brightness (although there are notable exceptions; see Section~\ref{sec:Perseus}), meaning that only the inner mass profile of a cluster would be reliably reconstructed using the aforementioned relation (Equation~\ref{eq:fit}). However, it is possible to extrapolate the ICL profile, so that the mass profile can be computed up to M$_{200}$. For this exploratory work, we decided to use a simple power-law extrapolation of the ICL slope.

The slope of the extrapolated power law was computed by fitting the outer region of the ICL profile between 50 kpc (500 kpc for Perseus) and the edge of the available data for each observation. We note that even from this simple extrapolation, important information about the quality of the reconstructed mass profile can be gained. Since the best fit has a logarithmic slope $a \sim -1.1$, we require the extrapolated ICL profile to have a slope of $n \lesssim -3.1$. Otherwise, the mass ($\propto \Sigma_\text{tot} (r) r^2 \propto \Sigma_\star r^{3.1}$) would remain unbounded, and the recovered virial masses and radii of the clusters would be unphysical.

\begin{figure}
  \resizebox{\hsize}{!}{\includegraphics{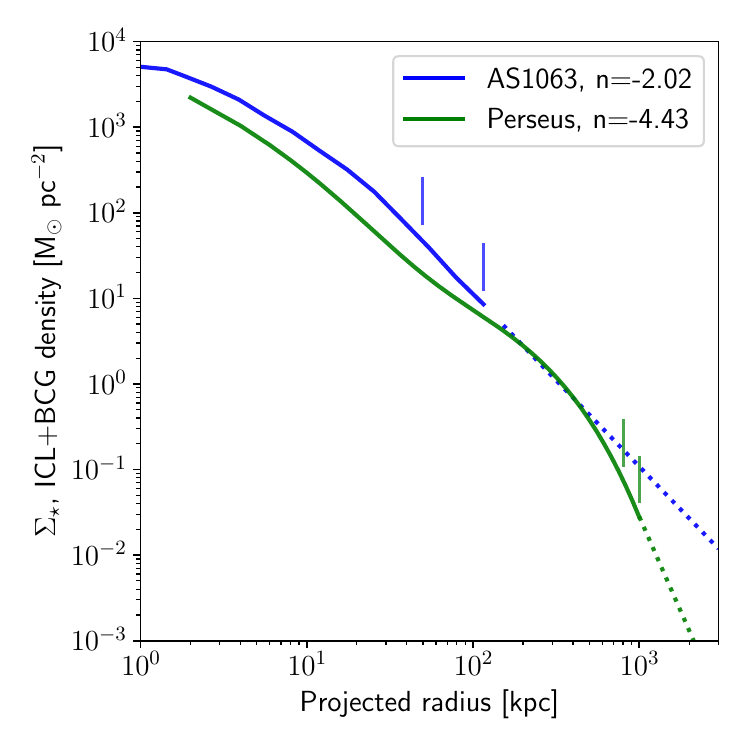}}
  \caption{Projected density profiles of the ICL+BCG for AS1063 \citep[blue,][]{Montes18} and Perseus \citep[green,][]{Kluge2025}. The profiles are extrapolated (dotted line) at large radii following the logarithmic slope in the regions marked with vertical lines. The observations of Perseus by Euclid reach the outskirts of the cluster, where $n \lessapprox -3$, and the extrapolated profiles do not produce unphysical massive clusters.}
  \label{fig:profiles}
\end{figure}

This is exemplified in Figure~\ref{fig:profiles}.
Two ICL+BCG density profiles are shown for Abell S1063 \citep{Montes18} and Perseus \citep{Kluge2025} clusters.
The former only reaches 200 kpc, where the logarithmic slope is $n=-2$.
The extrapolated profile for Abell S1063, as shown by the dotted lines, would result in a cluster with unbound mass.
On the other hand, the ICL profile of Perseus obtained with Euclid reaches a logarithmic slope of $n=-4.4$, so that the total matter profile is $\Sigma_\text{tot} \propto r^{-3.3}$ and the total mass of the cluster is finite.

To obtain error estimates for the mass profiles of the clusters, we performed a bootstrapping over the parameters of the $\Sigma_\star / \Sigma_\text{tot}$ fit to a multivariate normal distribution. We drew $10^4$ sample profiles for each cluster and built the median 1 and 2$\sigma$ percentiles at each radial distance.

\section{Results} \label{sec:results}

\subsection{Perseus cluster} \label{sec:Perseus}

To provide a first test of the aforementioned method, we used the exquisite data provided by the Euclid Early Release Observations of the Perseus cluster \citep{Kluge2025, Cuillandre2025}.
In this release, \citet{Kluge2025} provide ICL measurements up to about one-third of the virial radius of the cluster in two filters. The ICL profile was computed making use of isophotal and iso-density contours to take into account the ellipticity of the ICL, and careful processing to reach a limit surface brightness of 27 mag sec$^{-2}$.

We extracted the BCG+ICL radial surface brightness from the double S\'ersic fit of the $H_\text{E}$ filter. As in \citet{Kluge2025}, we derived the intracluster stellar mass density profile using Equation (1) of \citet{Montes2014} assuming a $M/L=1.02$ \citep{Vazdekis2016}.

\begin{figure}
  \resizebox{\hsize}{!}{\includegraphics{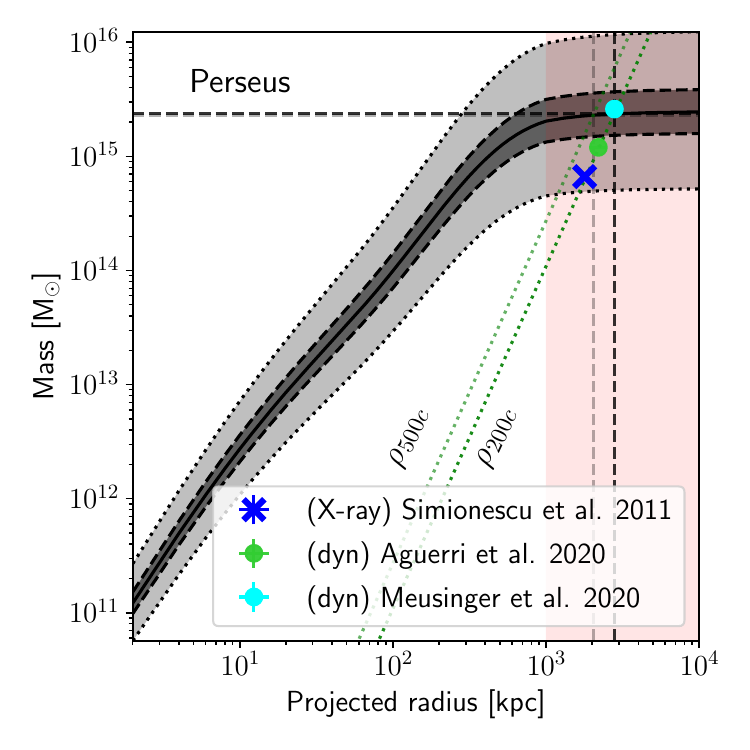}}
  \caption{Mass profile of the Perseus clusters as extracted from its ICL observed by Euclid \citep{Kluge2025}. The shaded regions represent the 1 and 2$\sigma$ intervals. The dashed black horizontal (vertical) lines indicate the recovered M$_{200c}$ (R$_{200c}$). The gray lines denote the same, but for M$_{500c}$. The red shaded regions indicate where $\Sigma_\star$ has been extrapolated. For reference, the locus where $\rho = \rho_{200c}$ is shown as dotted green lines.}
  \label{fig:perseus}
\end{figure}

After we obtained the intracluster stellar mass density profile of the Perseus cluster, the cluster mass profile was derived as described in Section~\ref{sec:methods}. The resulting profile is shown in Figure~\ref{fig:perseus}. The 1 and 2$\sigma$ confidence intervals are shown as gray shades. The obtained M$_{200}$ (R$_{200}$) are shown as dashed horizontal (vertical) black lines. The same is shown in gray for M$_{500}$ (R$_{500}$), although for Perseus both lines overlap due to the flatness of the mass profile at $\gtrsim 1$ Mpc.
The region where the intracluster stellar density profile has been extrapolated assuming a power-law is indicated with a red shade. For the Perseus cluster, we include mass estimates from galaxy velocity dispersion \citep{Aguerri2020, Meusinger2020} and X-ray \citep{Simionescu2011}. The recovered mass of Perseus from the ICL is $M_{200} = 2.4^{+1.3}_{-0.9} \,\times 10^{15}\,\text{M}_\odot$. The mass profile is mostly consistent with observations, especially with the cluster mass measure of \citet{Meusinger2020} computed from a sample of 286 spectroscopically confirmed galaxy members.

\subsection{Hubble Frontier Fields Clusters} \label{sec:HFF}

\begin{figure*}
  \resizebox{\hsize /2}{!}{\includegraphics{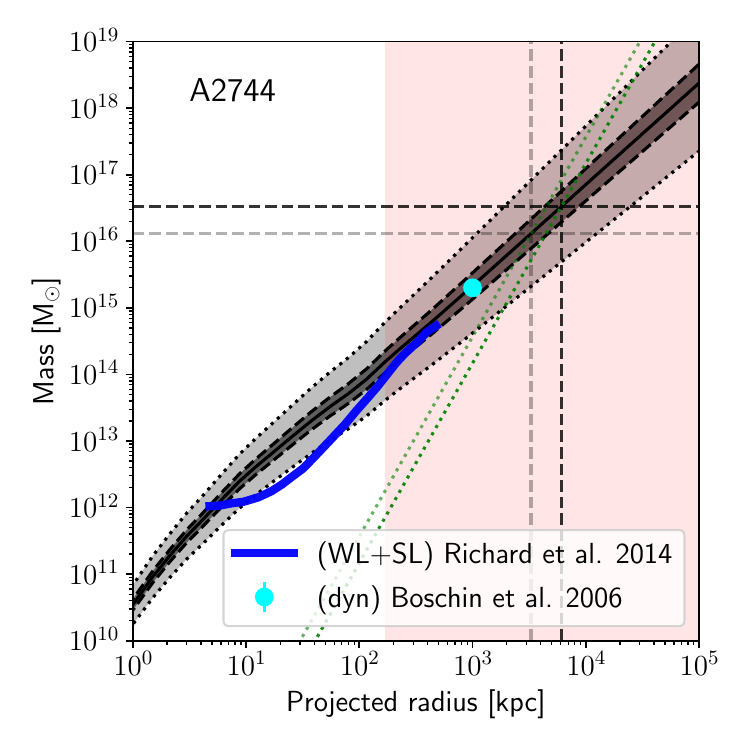}}
  \resizebox{\hsize /2}{!}{\includegraphics{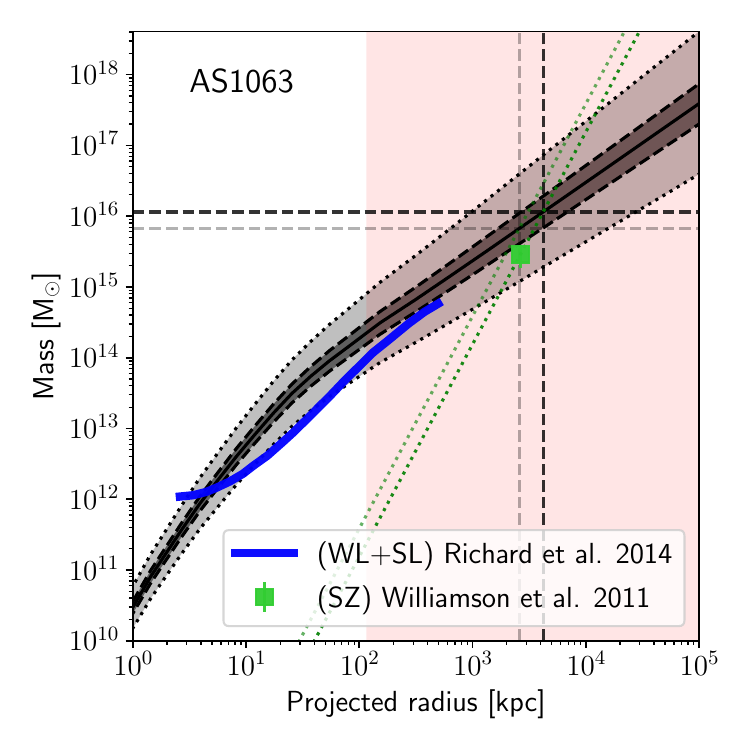}}
  \resizebox{\hsize /2}{!}{\includegraphics{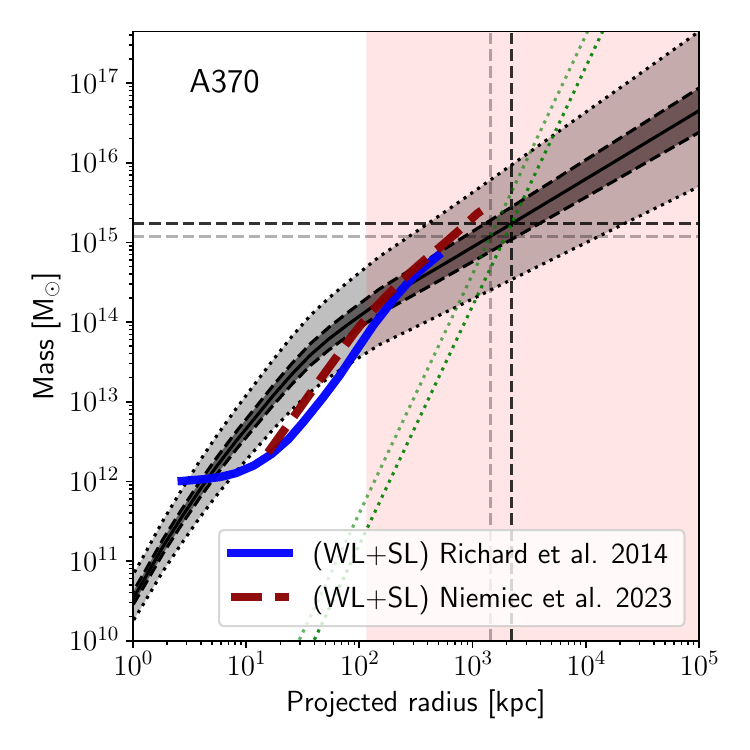}}
  \resizebox{\hsize /2}{!}{\includegraphics{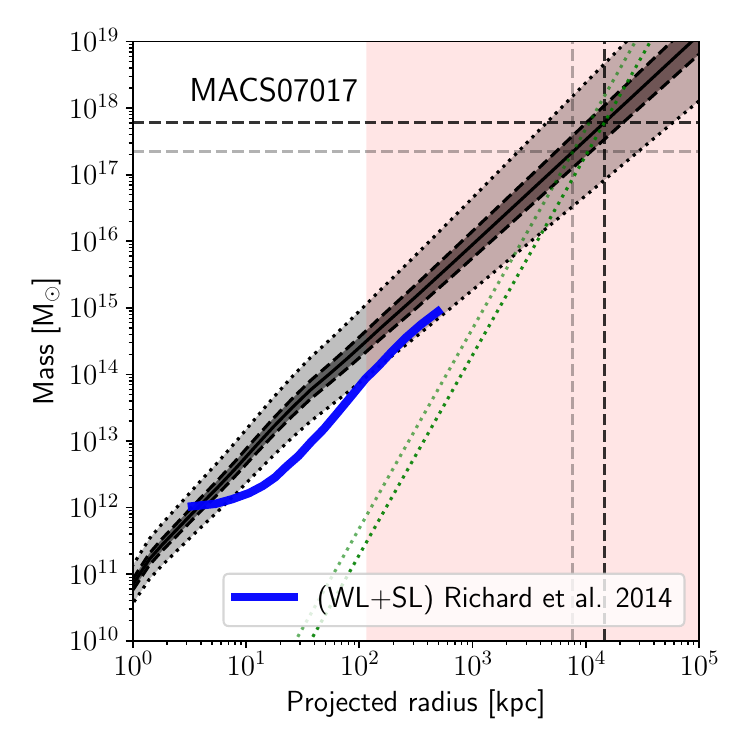}}

  \caption{Same as figure~\ref{fig:perseus}, but for the HFF clusters. Because the ICL observations do not reach large enough radii, the mass of the cluster remains unbound.}
  \label{fig:HFF}
\end{figure*}

We show in Figure~\ref{fig:HFF} the application of our method to four clusters that are part of the Hubble Frontier Fields (HFF) survey \citep{Lotz2017a}. The fraction of ICL light and surface density profiles of the ICL+BCG were extracted directly from \citet{Montes18}, Appendix C.
We show for reference the profiles resulting from the detailed strong- and weak-lensing analysis by \citet{Richard2014}.

In the upper left panel, we show the results for Abell 2744 ($z=0.308$, $f_\text{ICL}=7.7\pm3.1$). For further comparison, we show the dynamical mass estimate from \citet{Boschin2006}. Although the dynamical mass is measured in the extrapolated region of the mass profile extracted from the ICL, both show very good agreement. This indicates that the power-law extrapolation, albeit simple, reproduces well the $\Sigma_\star$ profile in the outer regions.

In the upper right panel, the mass profile for Abell S1063 ($z=0.348$, $f_\text{ICL}=13.1\pm2.8$) is shown together with M$_{200}$ computed from the Sunyaev-Zeldovich effect by \citet{Williamson2011}. In this case, the mass is overestimated, but observations lie close to the 1$\sigma$ region.

In the lower left panel, we show the results for Abell 370 ($z=0.375$, $f_\text{ICL}=4.8\pm1.7$) and an updated strong- plus weak-lensing mass profile from \citet{Niemiec2023}, which extends to larger radii than those of \citet{Richard2014}.
Namely, we show the baseline model (model$_\text{A}$) of \citet{Niemiec2023}.
They observed a flat density profile inside 100 kpc of $3 \times 10^9$ M$_\odot$ kpc$^{-2}$, which is compatible with other lensing analyses of A370 \citep{Lagattuta2019}.
The recovered mass profile of A370 from the ICL also shows a nearly flat density profile, but its density is overestimated to $10^{10}$ M$_\odot$ kpc$^{-2}$.
This produces the visible differences in the inner regions of the cluster.

In the lower right panel, the results for MACSJ0717.5+3745 ($z=0.545$, $f_\text{ICL}=5.7\pm1.6$) are shown.
From our sample, this is the farthest galaxy cluster and presents overall worse agreement with observations compared to the rest of the sample.
As we used the $\Sigma_\star / \Sigma_\text{tot}$ relation from simulations at lower redshift, this may indicate that for higher-redshift observations, the relation is not valid or should at least be recomputed.
This is likely caused by the dynamical state of the cluster and its ICL at higher redshift, as only half the ICL is assembled by redshift $z=0.5$ \citep[][Contreras-Santos in prep.]{Contini14}.
We give an overview of the time dependence of the fit parameters in Appendix~\ref{app:fit_parameters} .

\begin{table}
    \caption{Mass of the clusters computed from their ICL following Sect.~\ref{sec:methods}.}
    \label{tab:m200}
    \centering
    \renewcommand{\arraystretch}{1.5}
    \begin{tabular}{c|c}
        \hline\hline
        Cluster & $M_{200}$ (M$_\odot$) \\ \hline\hline
        Perseus & $2.4^{+1.3}_{-0.9} \,\times 10^{15}$ \\ \hline
        A2744 & $3.5^{+7.1}_{-2.3} \,\times 10^{16}$ \\
        AS1063 & $1.1^{+1.4}_{-0.6} \,\times 10^{16}$ \\
        A370 & $1.8^{+1.8}_{-0.9} \,\times 10^{15}$ \\
        MACS07017 & $5.8^{+14.1}_{-4.0} \,\times 10^{17}$ \\\hline
    \end{tabular}
    \tablefoot{Only for Perseus can the ICL profile be measured at large enough radii to provide an accurate mass estimation.}
\end{table}

We show in Table~\ref{tab:m200} a summary of the recovered M$_{200}$ in this and the previous sections.
For all HFF clusters, the recovered M$_{200}$ (and M$_{500}$) is generally unphysically large because the mass of the cluster remains unbounded. Regarding the mass profiles, the extrapolation shows good agreement with observations out to $\sim 1$ Mpc. It can be noted a general mass overestimation compared to lensing measurements of a factor $\sim 5$ at 20-30 kpc, that decreases at larger radii. But overall, the observations lie within the 2$\sigma$ region of the cluster mass profile recovered from the ICL.

\section{Discussion} \label{sec:discussion}

The results presented in the previous sections show that the ICL can be used to
measure the cluster mass profile using a simulation-calibrated relation.
Although the errors in the mass profiles are still large
($\pm \sim 0.5$ dex at 100 kpc and $\pm \sim 0.7$ dex at 1 Mpc),
this method provides an independent measure of the cluster mass, which can be
of value when the measures from different tracers differ (e.g., X-ray
and lensing).
The ICL can be typically well determined in the inner regions of the cluster
($\lesssim 300$ kpc), and in these cases, it can be used to trace the detailed
structure of the total mass of the cluster.

We note, however, that the method explored in this work has caveats and potential improvements:
\begin{enumerate}
    \item
      When the ICL profile does not reach a logarithmic slope of $\lesssim -3$,
      the total mass of the cluster will remain unbounded assuming a power-law
      extrapolation of the profile (Figure~\ref{fig:profiles}). If this
      assumption is relaxed, another extrapolation could be applied, but is
      likely to introduce some prior knowledge of the cluster mass distribution
      (e.g., M$_{200}$ and R$_{200}$). One option that would not include such
      priors would be to fit the available ICL profile to a double S\'ersic
      profile. \citet{Kluge2025} showed that it fits well the
      observed ICL profile up to $\sim 1$ Mpc. But even in this case, the
      profile should be measured to at least a few hundred kiloparsecs to have
      reliable parameters of the S\'ersic profiles.
    \item
      The relation $\Sigma_\star / \Sigma_\text{tot}$ was calibrated from
      simulations of the C-EAGLE project, which comprises 30 clusters in the mass
      range $[10^{14}, 10^{15.4}]$ M$_\odot$. Within this range, we find no
      significant correlation of the fit parameters with the cluster mass
      (Figure~\ref{fig:parameters}). However, it would be valuable to expand the
      analysis with other simulations to increase the statistical significance and
      study the impact of the galaxy formation model on this relation. It must be noted that these simulations need to be of high resolution to have a
      non-negligible population of stars belonging to the ICL in the outskirts of
      the galaxy cluster \citep{Martin2024a}.
    \item
      We did not explore the dependence on the cluster relaxation state,
      which can affect the amount of ICL and its distribution
      \citep[e.g.,][]{Contreras-Santos2024}, which in turn adds biases to the
      proposed mass estimation method we presented.
\end{enumerate}

As the simulation-calibrated $\Sigma_\star / \Sigma_\text{tot}$ relation
presented in this work has the aforementioned caveats, one can open the
question: \textit{can the relation between the total matter density profile and
the stellar mass density profile be derived directly form observations?} If
that would be the case, this analysis could be repeated without the need of any
theoretical model or assumption, improving significantly the impact and
applicability of the method laid out in this work.  However, this would require
a substantial observational effort and the applicability of such relation to
clusters for different masses, relaxation states and redshifts would still need to
be informed by theoretical studies \citep{Butler2025}.

\section{Conclusions} \label{sec:conclusions}

We proposed and tested a simulation-calibrated method for inferring the
mass profiles of clusters by measuring their ICL profiles.

By extending the relation established by \citet{AlonsoAsensio2020} between ICL
and total mass profiles, we refined the error estimates and provided
a fitting formula (Equation~\ref{eq:fit}) and the parameters
(Table~\ref{tab:fit_parameters}). In turn, they were used to obtain
an estimate of the cluster total mass profile.  The method was tested
on both ideal and suboptimal ICL observations. The application to high-quality
Euclid data of the Perseus cluster \citep{Kluge2025} showed the potential to recover a robust cluster mass profile that extends to the outskirts, constraining
the virial mass and radius of the cluster (Figure~\ref{fig:perseus}).

Our findings emphasize the importance of ICL observations reaching beyond
several hundred kiloparsecs, where the ICL logarithmic slope approaches -3.0 or
less (Figure~\ref{fig:profiles}). Otherwise, a simple power-law extrapolation
produces unphysically unbound clusters (Figure~\ref{fig:HFF}). Improvements to
the extrapolation technique could offer more precise mass profiles, but care
must be taken to avoid introducing biases on the mass or size of the cluster.
Further studies using higher-resolution simulations could broaden the parameter
space of cluster masses and redshifts covered, bolstering the method's generalizability. Additionally, the potential dependence on the cluster dynamical
state, redshift, and formation history warrants further exploration.

Our approach highlights the ICL's capability as a mass estimator for galaxy
clusters and presents an observationally accessible mass indicator that could
complement traditional methods such as weak lensing and X-ray.
This is especially relevant as Euclid will detect the ICL of several hundred
clusters at more than 500 kpc from the cluster center, up to $z=0.7$ \citep{EuclidCollaboration2025}.
Future observational efforts to directly calibrate the ICL-mass relation from a broad
sample of clusters would eliminate theoretical dependences and broaden the
applicability of this method.

\begin{acknowledgements}
  We are grateful to the anonymous referee, who has helped to improve the clarity of the manuscript.
  We thank Claudio Dalla Vecchia, Mireia Montes and Ignacio Trujillo for fruitful discussions.
  This work has been supported by the Spanish Ministry of Science, Innovation and Universities (\textit{Ministerio de Ciencia, Innovaci\'on y Universidades}, MICIU) through research grants PID2021-122603NB-C21 and PID2021-122603NB-C22.
  The author(s) acknowledge the contribution of the IAC High-Performance Computing support team and hardware facilities to the results of this research.
\end{acknowledgements}

\bibliographystyle{aa}
\bibliography{biblio}

\begin{appendix}

  \section{Fit parameters} \label{app:fit_parameters}

  In this Appendix we present the fit parameters for the different studied
  epochs: $z=0, 0.1, 0.2, 0.35, 0.54$ and an aggregate of all the data with $z<0.5$.

  All the fits were performed to a multivariate normal distribution defined by the probability density distribution
  \begin{equation}
    f(x) = \frac{1}{\sqrt{(2 \pi)^2 \det \Sigma}}
                   \exp\left( -\frac{1}{2} (x - \mu)^T \Sigma^{-1} (x - \mu) \right),
  \end{equation}
  where $\mu=(\bar{a}, \bar{b})$ is the mean values of the power law fit
  (Equation~\ref{eq:fit}). $\Sigma$ is the covariance matrix. Its diagonal members are
  the variances $\Sigma_a$ and $\Sigma_b$, and the off-diagonal value
  $\Sigma_{ab}=\Sigma_{ba}$ is the covariance.

  \begin{table}\centering
    \caption{Fit parameters to a multivariate normal distribution for different epochs. \label{tab:fit_parameters}}
    \begin{tabular}{ c|c c c c c }
      Dataset  & $\bar{a}$    & $\bar{b}$   & $\Sigma_a$  & $\Sigma_b$  & $\Sigma_{ab}$ \\ \hline
      $z=0$    & -1.118 & 0.329 & 0.00896     & 0.047       & -0.0171       \\
      $z=0.1$  & -1.127 & 0.305 & 0.00928     & 0.033       & -0.0136       \\
      $z=0.2$  & -1.147 & 0.321 & 0.01076     & 0.038       & -0.0169       \\
      $z=0.35$ & -1.162 & 0.309 & 0.00852     & 0.029       & -0.0119       \\
      $z=0.54$ & -1.123 & 0.165 & 0.01545     & 0.066       & -0.0263       \\
      $z<0.5$  & -1.139 & 0.316 & 0.00958     & 0.036       & -0.0146       \\
    \end{tabular}
  \end{table}

\begin{figure}
  \resizebox{\hsize}{!}{\includegraphics{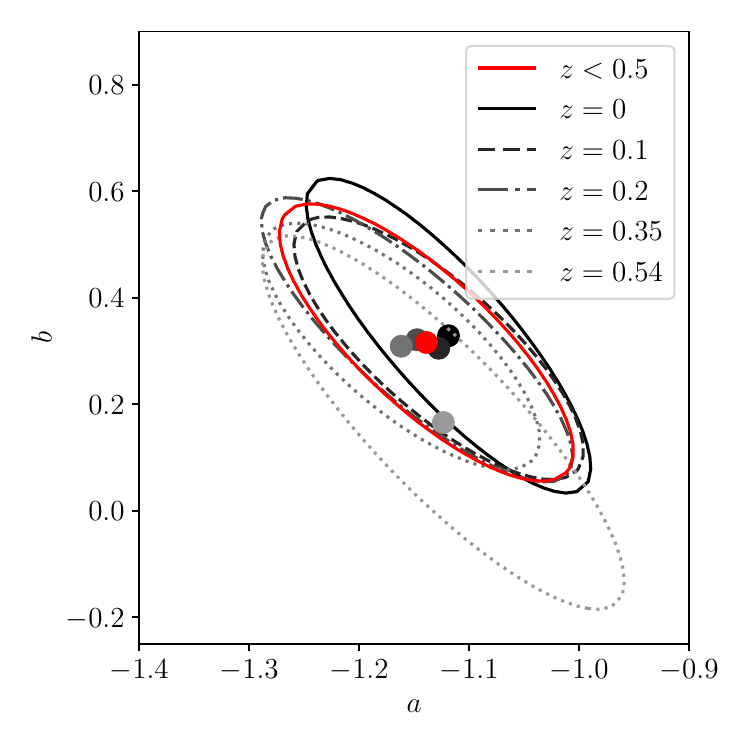}}
  \caption{1-$\sigma$ contours of the fit parameters of equation \ref{eq:fit} for different redshift. The aggregated dataset with $z<0.5$ is shown in red.}
  \label{fig:parameters_compare_z}
\end{figure}

  The fit parameters (i.e., $\bar{a}$, $\bar{b}$, $\Sigma_a$, $\Sigma_b$ and
  $\Sigma_{ab}$) are shown in Table~\ref{tab:fit_parameters}.
  To better illustrate the redshift evolution of the parameters, we show in
  Fig.~\ref{fig:parameters_compare_z} the mean and 1-$\sigma$ contours for each epoch.
  The dataset with $z<0.5$ is shown in red, whereas each epoch
  is shown in greyscale, darker shades indicate more recent epochs.

  For all $z<0.5$, there is no noticeable redshift evolution of the fits, indicating that the ICL has been assembled and is relaxed
  in the cluster potential at this point. However, for $z=0.54$, there is a noticeable wider range of possible parameters.
  This indicates that the ICL may not be fully relaxed by that time.
  This view is compatible with other works that show that only half the ICL is assembled by redshift $z=0.5$ \citep[][Contreras-Santos in prep.]{Contini14}.

\section{Mass dependence} \label{app:mass}

\begin{figure}
  \resizebox{\hsize}{!}{\includegraphics{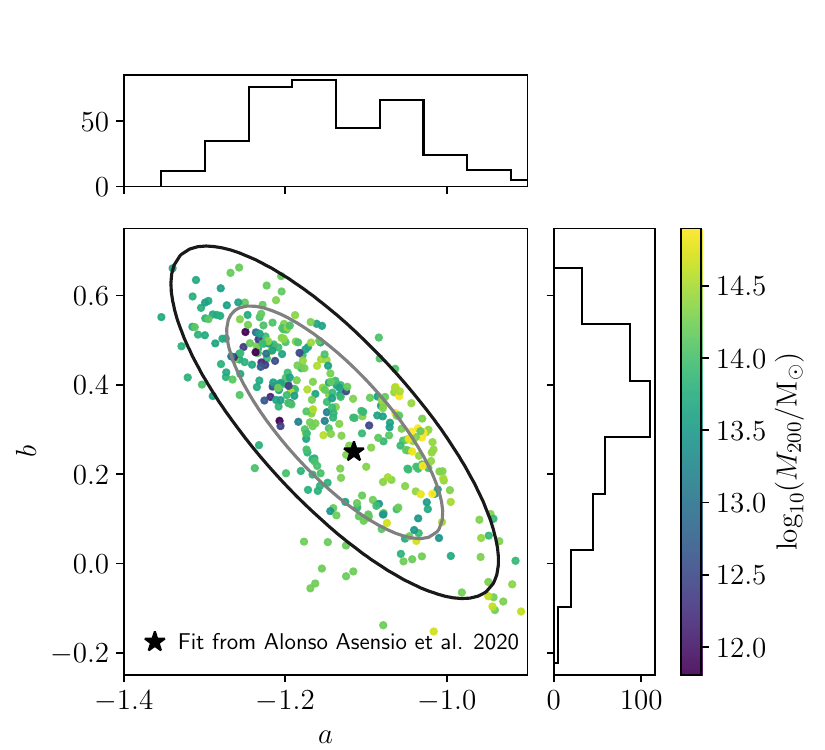}}
  \caption{Corner plot of the fit parameters of equation \ref{eq:fit} for $z<0.5$. The contours show the 1-$\sigma$ regions of the multivariate normal distribution of the fits. The fit parameters are shown in Table~\ref{tab:fit_parameters}}
  \label{fig:parameters_all_z}
\end{figure}

  To study the possible mass dependence of the fit parameters we show them in Figure~\ref{fig:parameters_all_z}, color-coded by the
  mass of the cluster. Overall, the distribution is well represented by the multivariate normal distribution, and there is only slight
  dependence of the mass, but always within the 1-$\sigma$ region.
  We have experimented to perform fits for different mass ranges, but found
  that due to the limited dataset the recovered fits were not accurate.
  Furthermore, we believe that the applicability of such fits may be limited as
  they require prior knowledge of the mass of the cluster.

\end{appendix}

\label{LastPage}
\end{document}